\documentclass[ aip,apl, amsmath,amssymb, reprint]{revtex4-1}

\usepackage{textcomp,color}
\usepackage[colorlinks=true,linkcolor=blue]{hyperref}
\usepackage{graphicx}
\usepackage{dcolumn}
\usepackage{bm}

\begin{document}

\title[Sample Title]{Tight-binding terahertz plasmons in chemical vapor deposited graphene}

\author{Andrey Bylinkin}
\affiliation{ 
Moscow Institute of Physics and Technology, Dolgoprudny 141700, Russia
}%

\author{Elena Titova}
\affiliation{ 
Moscow Institute of Physics and Technology, Dolgoprudny 141700, Russia
}%

\author{Vitaly Mikheev}
\affiliation{ 
Moscow Institute of Physics and Technology, Dolgoprudny 141700, Russia
}%

\author{Elena Zhukova}%
\affiliation{ 
Moscow Institute of Physics and Technology, Dolgoprudny 141700, Russia
}

\author{Sergey Zhukov}%
\affiliation{ 
Moscow Institute of Physics and Technology, Dolgoprudny 141700, Russia
}

\author{Mikhail Belyanchikov}
\affiliation{ 
Moscow Institute of Physics and Technology, Dolgoprudny 141700, Russia
}

\author{Mikhail Kashchenko}
\affiliation{ 
Moscow Institute of Physics and Technology, Dolgoprudny 141700, Russia
}

\author{Andrey Miakonkikh}
\affiliation{ 
Valiev Institute of Physics and Technology of Russian Academy of Sciences, Moscow 117218, Russia
}%

\author{Dmitry Svintsov}
\affiliation{ 
Moscow Institute of Physics and Technology, Dolgoprudny 141700, Russia
}%

\begin{abstract}
Transistor structures comprising graphene and sub-wavelength metal gratings hold a great promise for plasmon-enhanced terahertz detection. Despite considerable theoretical effort, little experimental evidence for terahertz plasmons in such structures was found so far. Here, we report an experimental study of plasmons in graphene-insulator-grating structures using Fourier transform spectroscopy in 5-10 THz range. The plasmon resonance is clearly visible above the Drude absorption background even in chemical vapor deposited (CVD) graphene with low carrier mobility $\sim 10^3$ cm$^2$/(V s). We argue that plasmon lifetime is weakly sensistive to scattering by grain boundaries and macoscopic defects which limits the mobility of CVD samples. Upon placing the grating in close proximity to graphene, the plasmon field becomes tightly bound below the metal stripes, while the resonant frequency is determined by the stripe width but not by grating period. Our results open the prospects of large-area commercially available graphene for resonant terahertz detectors.
\end{abstract}

\maketitle

Graphene-based optoelectronic devices benefit from high-speed operation~\cite{koppens2014photodetectors,ultrafast_graphene}, broadband response~\cite{broadband_graphene}, and compatibility with on-chip optical interconnects~\cite{chip_integrated_graphene}. Their major drawback is low electromagnetic wave absorbance by a single sheet of graphene. This problem is readily resolved via coupling of incident light to plasmons bound either to adjacent metal nanoparticles~\cite{plasmon_multicolor,plasmonic_enhancement_photovoltage} or to graphene itself~\cite{freitag_photodetector_intrinsic_plasmons}. Unlike plasmons in metals, intrinsic graphene plasmons offer ultra-strong field confinement~\cite{Koppens_confinement_limits} and tuning of resonant frequency with gate voltage~\cite{ryzhii2007plasmon,fei_gate_tuning}.

Resonant excitation of plasmons in graphene-based photodetectors becomes increasingly difficult when going from infrared to terahertz (THz) range~\cite{nikitin_acoustic_thz} as the plasmon quality factor scales linearly with frequency. Despite considerable effort~\cite{vicarelli_thz,Olbrich_Ratchet,Bandurin_dual_origin} evidence of plasmon-assisted THz detection in graphene are scarce and were reported only for high-quality encapsulated graphene~\cite{bandurin2018resonant} or epitaxial graphene on SiC~\cite{cai2015plasmon}. Experimental demonstrations of terahertz plasmons in absorbance spectra of graphene, including scalable chemical vapor deposited (CVD) samples, are more numerous~\cite{ju_plasmons_ribbons,gao_excitation_active_control,THz_plasmons_graphene_insulator_stacks}. At the same time, most such experiments dealt with ribbon-patterned where collection of photocurrent is hindered and boundary scattering is enhanced.

\begin{figure}
\includegraphics[width=1\linewidth]{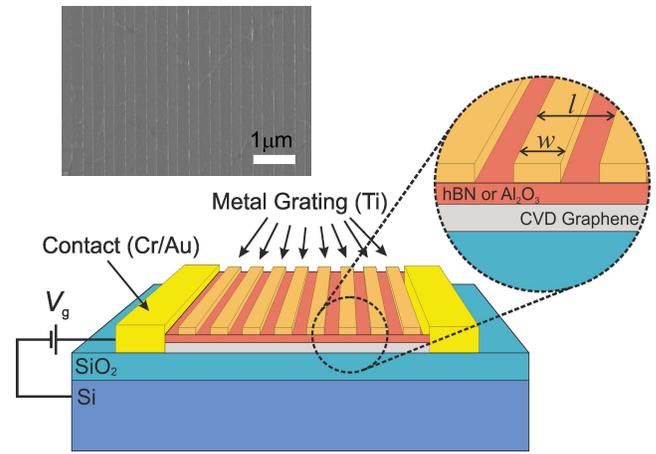}
\caption{\label{fig:fig1a} 
Schematic view of graphene-based transistor with sub-wavelength grating for launching of THz plasmons. Gate voltage applied between source and Si substrate controls the carrier density. Inset: SEM image of metal grating of device \#3 
}
\end{figure}

In this paper, we study the plasmonic properties of a basic building block of graphene-based terahertz detector~\cite{Olbrich_Ratchet,Fateev_rectification}, the CVD graphene-channel field-effect transistor with a grating gate. We find that plasmonic contribution to absorption spectra is pronounced at $5-10$ THz frequencies despite moderate carrier mobility $\sim 10^3$ cm$^2$/V s and short momentum relaxation time $\tau_{p} \sim 50$ fs. We further argue that plasmon lifetime in CVD-graphene (as it enters the quality factor) exceeds the relaxation time as extracted from mobility, in contrast to reports for encapsulated graphene. We find that metal grating placed in immediate vicinity to graphene modifies the resonant plasmon frequencies. In particular, the reciprocal wave vector of the grating no more determines the discrete frequencies of plasmons (contrary to the pioneering studies~\cite{Allen_Tsui_Logan,Mikhailov_Instability,Mackens_minigaps,Chaplik_absorption_emission}). Instead, the length of metal fingers plays the role of plasmon ''quantization length'', and the plasmon field is tightly bound below the metal gratings.

Our devices schematically shown in Fig.~\ref{fig:fig1a} were made of commercially available CVD graphene (Graphene Supermarket). Large-area ($\sim 5\times 5$ mm) graphene films were wet-transferred on oxidized silicon substrates using the established techniques~\cite{suk2011transfer}. The resistivity of Si substrate, $\sim 10-20$ Ohm$\cdot$cm, was low enough to achieve efficient gating of graphene, but still large enough to ensure transparency to the incident THz radiation, $\sim 40$ \% in the frequency range  50 -- 650 cm$^{-1}$. The top dielectric layers were made from CVD hexagonal boron nitride (hBN) or atomic layer-deposited Al$_2$O$_3$. Use of dielectrics with different permittivity and thickness enabled us to demonstrate plasmon resonance across the whole $5-10$ THz range. Atop the dielectric layer, we fabricated deep-subwavelength metal gratings [shown in \ref{fig:fig1a}, inset] with fixed period $l=500$ nm and different filling factors. The grating generates evanescent fields with large in-plane wave vectors and launches of graphene plasmons. The structural parameters of devices are summarized in Table \ref{tab:table1}.

\begin{figure}
\includegraphics[width=1\linewidth]{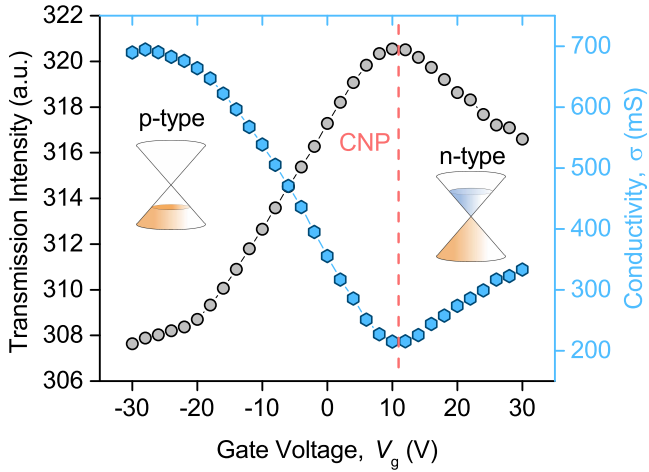}
\caption{\label{fig:fig1b} 
Spectrally-integrated transmission intensity and two-terminal graphene conductivity vs gate voltage for device \# 1. Anti-correlation between transmission and conductivity confirms that gate-dependent part of absorption is due to graphene
}
\end{figure}

The gate-controlled THz transmission spectra were measured using Fourier spectrometer Bruker Vertex 80v with mercury lamp as a source and silicon bolometer as a detector. Simultaneously with THz measurements, we controlled the two-terminal dc conductivity of the devices using with Keithley 2612B source-meter. The spectrally-averaged intensity of transmitted radiation [shown with grey dots in Fig.~\ref{fig:fig1b}] demonstrates a pronounced anti-correlation with channel conductivity. This proves that gate-dependent part of spectra is governed by graphene channel but not by, e.g., inversion layer of Si substrate.

Measurements of gate-controlled dc conductivity shown in Fig.~\ref{fig:fig1b} enable the estimate of field-effect mobility $\mu$. To exclude the contribution of contact resistance $R_{\rm cont} $, we have fitted the $R(V_{\rm g})$ dependence as
\begin{equation}
R(V_{\rm g}) = R_{\rm cont} + \frac{L/W}{\mu C |V_{\rm g}-V_{\rm cnp}|},   
\end{equation}
where is the resistance of graphene channel of length $L$ and width $W$, $C$ is the gate-to-channel capacitance, $V_{\rm g}$ is the gate voltage and $V_{\rm cnp}$ is the voltage corresponding to charge neutrality point. The extracted mobility was in the range $(1.0 - 1.5)\times 10^3$ cm$^2$/ V s, typical for wet-transferred CVD graphene \cite{buron2015graphene}. This value translates into carrier momentum relaxation time $\tau_{\rm dc} = \varepsilon_F \mu/e v_0^2$ in the range $20-50$ fs ($\varepsilon_F$ is the gate-controlled Fermi energy and $v_0=10^6$ m/s is the Fermi velocity). Such short relaxation time precludes the emergence of terahertz plasmon resonance which we nonetheless attempted to observe.

\begin{table}
\caption{\label{tab:table1} Grating parameters of studied devices. $w$ is the metal stripe width, $l$ is the grating period, $d$ is the insulator thickness}
\begin{ruledtabular}
\begin{tabular}{ccccc}
\mbox{ Sample \# } & \mbox{ $w$, nm} & \mbox{ $l$, nm} &\mbox{ $d$, nm}& \mbox{Insulator}\\
\hline
1 & 170 & 500 & 11 & Al$_2$O$_3$ \\
2 & 170 & 500 & 6 & Al$_2$O$_3$\\
3 & 300 & 500 & 14 & hBN\\
\end{tabular}
\end{ruledtabular}
\end{table}

Extraction of plasmonic contribution to the terahertz absorption is a non-trivial task due to residual free-carrier (Drude) absorption in Si substrate and graphene itself. The substrate contribution is eliminated by dividing the transmitted spectrum $T(V_{\rm g})$ by the one recorded at the charge neutrality point, $T(V_{\rm cnp})$. The normalized spectra, $T(V_{\rm g})/T(V_{\rm cnp})$, carry essentially different information depending on the polarization of probe light~\cite{cai2015plasmon}.

\begin{figure*}
\includegraphics[width=1\linewidth]{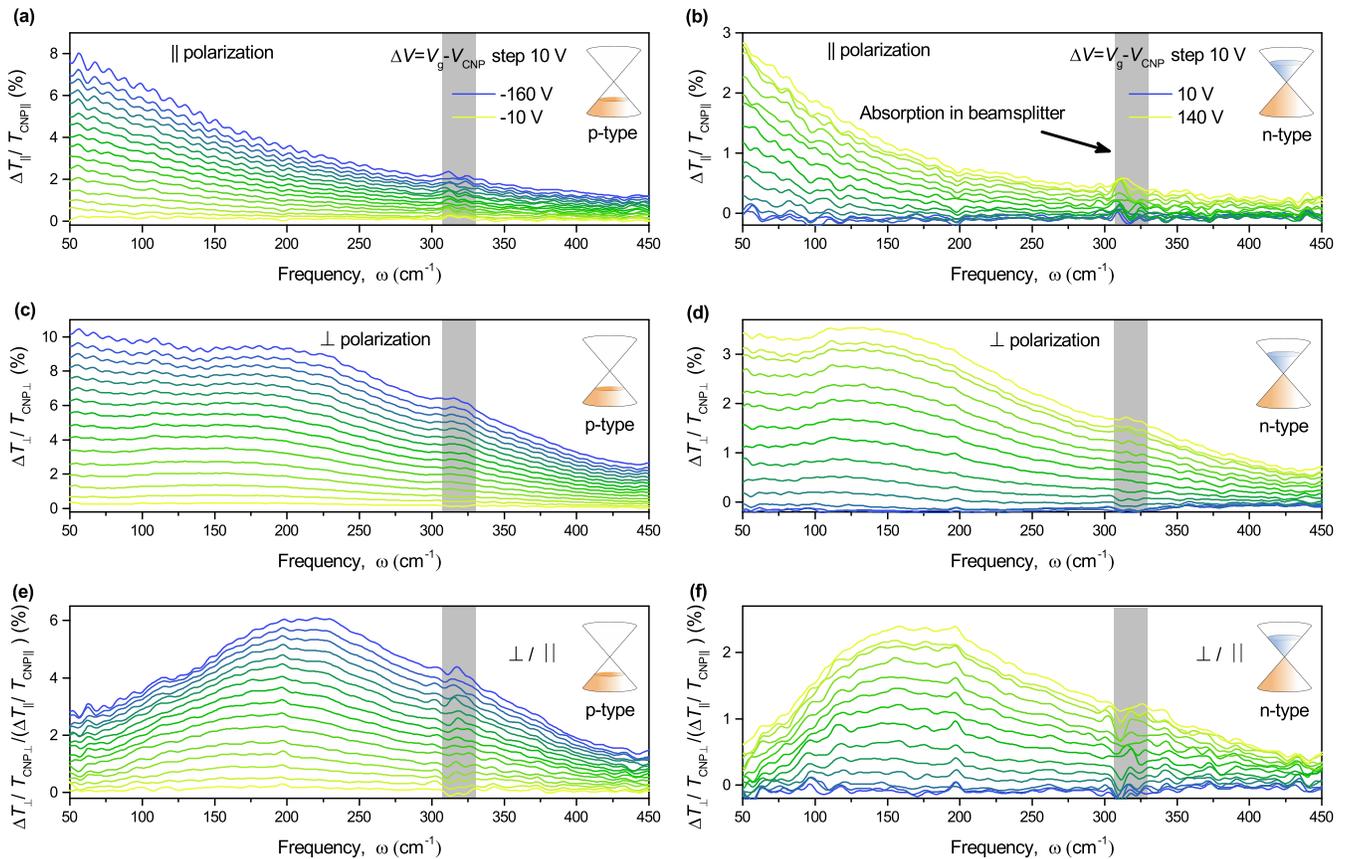}
\caption{\label{fig:fig2}
Transmission spectra of graphene-based transistor with metal grating measured at two mutually orthogonal polarizations and different gate voltages. Left column corresponds to hole doping, right column -- to electron doping. Top line: $\bf E$-field polarized along the gratings; middle line: $\bf E$-field polarized transverse to the gratings. Bottom line: plasmonic contribution to absorbance obtained by dividing the 'transverse spectra' by 'parallel' ones.
}
\end{figure*}

Light with ${\bf E}$-field parallel to metal gratings cannot efficiently generate evanescent waves, and the normalized transmittance is almost the same as of uniform  graphene film with conductivity $\sigma(V_{\rm g})$ on a substrate with refractive index $n_{\rm Si}$:
\begin{equation}
\label{e:FitDrudeTransmission}
\left[\frac{ T (V_{\rm g}) }{ T(V_{\rm cnp})}\right]_\parallel \approx  1 - \frac{8\pi/c}{1+n_{\rm Si}}[\sigma(V_{\rm g})-\sigma(V_{\rm cnp})].
\end{equation}
The absorption spectra measured in such polarization, $[T(V_{\rm g})/T(V_{\rm cnp})]_\parallel$ are shown in Figs.~\ref{fig:fig2}(a) and \ref{fig:fig2}(b) for hole and electron doping, respectively. In accordance with the above, they possess no resonant features and follow the smooth trend of frequency-dependent graphene conductivity (see the discussion below).

Light with ${\bf E}$-field transverse to metal gratings is strongly converted to near field with large in-plane wave vector and couples to plasmons. As a result, the spectra recorded in transverse polarization [shown in \ref{fig:fig2}(c-d)] possess an additional feature with maximum shifting toward higher frequencies with increasing carrier density. Further quantitative analysis confirms its plasmonic origin. 

Already the quantitative analysis of absorption in ''parallel'' polarization provides 
valuable information about high-frequency transport properties of graphene, complementary to the information from dc measurements. In particular, fitting of these spectra with Drude model enables simultaneous determination of carrier density and scattering rates. With account of both electrons and holes in graphene, the Drude conductivity takes on the form
\begin{equation}
\label{e:ConductivityGraphene}
\sigma(V_{\rm g})=\frac{e^2}{\pi \hbar} \frac{ i |\varepsilon_F (V_{\rm g})|/\hbar}{\omega + i \gamma },
\end{equation}
where $\gamma$ is the carrier scattering rate. 

\begin{figure}
\includegraphics[width=1\linewidth]{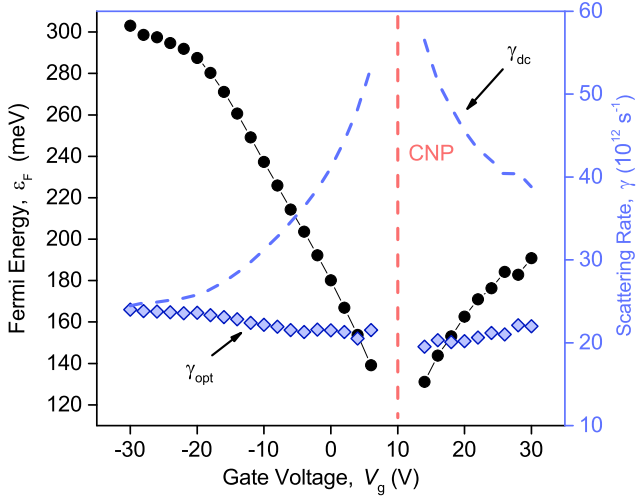}
\caption{\label{fig:fig3a}
Transport properties of graphene samples deduced from optical absorption data: Fermi energy vs gate voltage (black) and scattering rate (blue) for device \# 2. Non-zero $\varepsilon_F$ at the charge neutrality point is due to impurity-induced fluctuations; its magnitude was estimated with theory of Ref.~\onlinecite{Adam_self-consistent}  
}
\end{figure}

The outcome of the fitting procedure is shown in Figure \ref{fig:fig3a}. Its most remarkable feature is that scattering rate extracted from optical measurements, $\gamma_{\rm opt}$, is well below the scattering rate estimated from mobility, $\gamma_{dc} = e v_0^2/(\mu |\varepsilon_{\rm F}|)$. While being unexpected for uniform samples, the difference between these quantities can be explained with account of the poly-crystaline structure of CVD-grown samples that represent a network of randomly-oriented grains with average size $l_{\rm grain}\sim 5$ $\mu$m. In dc measurements, one determines the net resistance of poly-crystaline network, and the value of resistance is limited by worst conducting elements. In optical measurements, the conductivities of individual parts of sample are summed up, and the poorly conducting parts do not contribute much to the net absorbance~\cite{Cervetti_single_grain_mobility}. We can further argue that $Q$-factor of plasmons is determined by optical scattering rate, as both the plasmon wavelength ($\lambda_{pl}\sim 500$ nm) and free path are below the size of the grain.  

\begin{figure}
\includegraphics[width=1\linewidth]{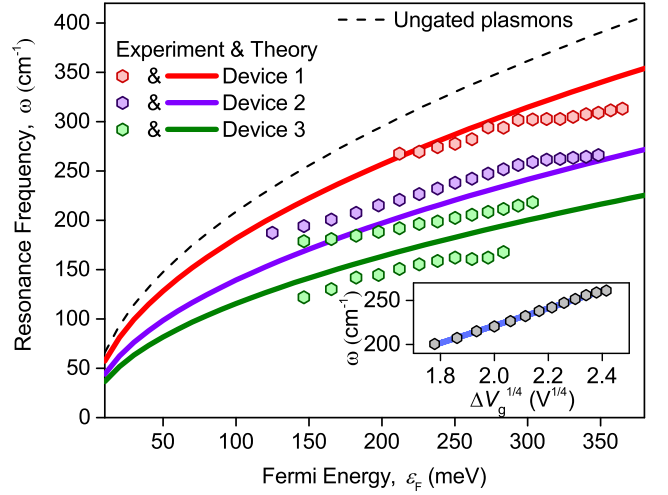}
\caption{\label{fig:fig3b}
 Extracted resonant frequencies of plasmons (dots) compared with tight-binding theory [Eq.~\ref{e:resonant_frequency}] for three samples. The lower and upper groups of points for sample \# 3 correspond to electron and hole doping, respectively. The black dashed line is the spectrum of ungated plasmons evaluated at reciprocal grating wave vector $G_{n=1} = 2\pi/l$. Inset: gate voltage dependence of resonant frequency $\Omega \propto |V_{\rm g}-V_{\rm cnp}|^{1/4}$
}
\end{figure}

The spectra recorded in transverse polarization, $[T(V_{\rm g})/T(V_{\rm cnp})]_\bot$, possess extra absorption peaks above the smooth Drude background. Their quantitative analysis is most conveniently performed after division of 'transverse' and 'parallel' transmittances. After such procedure, the Drude contribution is largely removed. Fitting the normalized spectra with  damped oscillator model ${\rm Im} \left[-\omega/\left(\omega(\omega + i\gamma)-\Omega^2\right)\right]$, we determined the resonant frequency $\Omega$~\cite{ju_plasmons_ribbons}. The density dependence $\Omega \propto n^{1/4}$ (inset in Fig.~\ref{fig:fig3b}) completes the proof that the feature is associated with graphene plasmon~\cite{ryzhii2007plasmon}. The values of scattering rate from damped oscillator model fit closely matched the scattering rate from Drude absorption, which further confirms that plasmon damping is weakly sensitive to grain boundaries and macroscopic defects.

The density dependence of plasmon resonance in graphene $\Omega \propto n^{1/4}$ is universal and independent of structure geometry~\cite{ryzhii2007plasmon}. Practical design of plasmonic THz detectors requires, however, the knowledge of relation between resonant frequency and structural parameters of the devices. It has been long assumed~\cite{Allen_Tsui_Logan,Mackens_minigaps,Mikhailov_Instability,Chaplik_absorption_emission} that resonant frequencies $\Omega_n$ of grating-coupled plasmons in 2d electron systems (2DES) are the frequencies of ungated plasmons $\omega_u(q)$ evaluated at reciprocal grating wave vectors $q=G_n = 2\pi n/l$. It implies that grating does not perturb the plasmon spectra of continuous 2DES, it only provides extra in-plane wave vector $G_n$ to the diffracted light. For massless carriers in graphene, these frequencies would be given by~\cite{ryzhii2007plasmon}
\begin{equation}
\label{Ungated}
   \Omega_n = \omega_{u}(G_n) = v_0 \sqrt{2 \alpha_c G_n k_F},
\end{equation} 
where $\alpha_c = e^2/\epsilon\hbar v_0$ is the Coulomb coupling constant, and $k_F = |\varepsilon_F|/\hbar v_0$ is the Fermi wave vector.

The inconsistency of the ''canonical'' picture with our data is apparent from Fig.~\ref{fig:fig3b}. The measured resonant frequencies for all samples lie below the frequency of ungated plasmon $\omega_{u}(G_{n=1})$. Moreover, samples with different metal filling factor $f=w/l$ and equal grating period have essentially different resonant frequencies, the sample with smallest $f$ having the highest frequency. This fact is not captured by the existing theory (\ref{Ungated}).

The main assumption of the previously accepted theories is weak coupling between grating and 2d plasmons. The coupling strength of the $n$-th mode is characterized by the quantity $e^{-G_n d}$ which is nothing but attenuation of $n$-th evanescent diffracted field at distance $d$ from the grating. In our devices, the coupling strength is order of unity, and assumptions of weak-coupling approach are not fulfilled. More generally, they would not be fulfilled in any sensitive plasmonic THz detector, because efficient conversion of light into plasmons requires weak attenuation of near-field harmonics.

We now present a simple quantitative model to estimate the resonant frequencies that is justified by electromagnetic modelling~\cite{popov2015noncentrosymmetric} and common sense, but its strict derivation is still missing. We model grating-coupled graphene as a sequence of ungated and gated regions with plasmon dispersions $\omega_u(q)$ and $\omega_g(q)$. The reflection coefficient from gated/ungated line boundary $r_{g/u}$ can be estimated be matching the electric potentials and currents~\cite{Petrov_Aplified_reflection}
\begin{equation}
    r_{g/u} = \frac{1-2q_g d}{1+2q_g d} \approx 1 - 4\pi\frac{d}{w}. 
\end{equation}
The latter equation implies that reflection of plasmon from gated/ungated boundary is almost perfect, up to the small corrections $\sim d/w$~\cite{Sydoruk_Reflection}. Nearly perfect reflection can be also interpreted with Fresnel's relation between reflection and phase velocities in adjacent media.

The considered situation is analogous to the tight-binding model of solids where band electrons are localized near individual atoms and hopping amplitudes are small. Here, the electric field of plasmon is concentrated below each individual metal stripe, and the probability of hopping to the neighbouring stripes is order of $d/w$. The length of metal therefore determines the discrete wave vectors of observed plasmons $q_n = \pi n/w$, while their dispersion is that of gated plasmon. As a result, the resonant frequencies should obey
\begin{equation}
\label{e:resonant_frequency}
    \Omega_n = \omega_g(q_n) = v_0 \frac{\pi n}{w} \sqrt{4 \alpha_c k_F d}.
\end{equation}

Without any fitting parameters, the predicted 'tight-binding' frequencies are close to those observed experimentally, as shown in Fig.~\ref{fig:fig3b}. In accordance with our model, reduction of filling factor while keeping the full period constant raises the resonant frequency, as observed for devices \#2 and \#3. Reduction in dielectric thickness leads to enhanced screening of Coulomb interaction and softening of plasmon modes, as seen from comparison of curves \#1 and \#2. 

Full test of the plasmon tight-binding model in grating-coupled graphene would require the observation of highest-order modes ($n>1$) in the transmission spectra. In the current setup, these effects are masked by plasmon hybridization with polar phonons in SiO$_2$ and hBN. Reduction of resonant frequency may resolve the situation, but would require samples of higher electronic quality.

The tight-binding character of plasmons in 2d systems placed in close proximity to gratings may have strong implications on design of resonant detectors. The fact that electric fields are concentrated below individual metal fingers implies the secondary role of grating periodicity for excitation of plasmons. They may be resonantly excited in aperiodic and even disordered gratings with fixed quantization length without appreciable broadening of resonant line. 

In conclusion, we have studied the properties of plasmons in large-scale chemical vapor deposited graphene with sub-wavelength metal grating. Possibility of THz plasmon excitation despite moderate carrier mobility shows the prospect of this structure as a voltage-tunable radiation detector. We have shown that the quality factor of plasmons is determined by single-grain optical mobility but not by dc mobility that is reduced due to macroscopic cracks and grain boundaries. For the practically important device design with grating in a close proximity to graphene, the resonant frequencies of plasmons are determined by the length of metal grating elements, while the electric fields are tightly bound to individual metal gratings.

The work was supported by the grant \# 16-19-10557 of the Russian Science Foundation. Samples were fabricated using the equipment of MIPT Shared Facilities Center with financial support from the Ministry of Education and Science of the Russian Federation (Grant No. RFMEFI59417X0014). The authors are grateful to V. Kaydashev, E. Korostylev, M. Zhuk and G. Fedorov for assistance.

\nocite{*}
\bibliography{aipsamp}

\end{document}